\documentclass[preprint,aps,showpacs,nofootinbib,preprintnumbers,amsmath,amssymb]{revtex4-1}
\usepackage{}
\usepackage{epsfig}
\usepackage{subfigure}
\usepackage{dcolumn}
\usepackage{bm}
\usepackage[usenames ,dvipsnames]{xcolor}
\usepackage{slashed}
\usepackage{graphicx,color}

\begin{document}
\title{Study of $\Lambda_b\to \Lambda (\phi,\eta^{(\prime)})$ and $\Lambda_b\to \Lambda K^+K^-$ decays}

\author{C.~Q. Geng$^{1,2,3}$, Y.~K. Hsiao$^{1,2,3}$, Yu-Heng Lin$^{3}$, and Yao Yu$^{1}$}
\affiliation{
$^1$Chongqing University of Posts \& Telecommunications, Chongqing, 400065, China\\
$^2$Physics Division, National Center for Theoretical Sciences, Hsinchu, Taiwan 300\\
$^3$Department of Physics, National Tsing Hua University, Hsinchu, Taiwan 300
}
\date{\today}

\begin{abstract}
We study the charmless two-body $\Lambda_b\to \Lambda (\phi,\eta^{(\prime)})$ and  three-body $\Lambda_b\to \Lambda K^+K^- $ 
decays. We obtain  ${\cal B}(\Lambda_b\to \Lambda\phi)=(3.53\pm 0.24)\times 10^{-6}$ to agree with the recent LHCb measurement.
However, we find that ${\cal B}(\Lambda_b\to \Lambda(\phi\to)K^+ K^-)=(1.71\pm 0.12)\times 10^{-6}$ is unable to explain the LHCb observation of ${\cal B}(\Lambda_b\to\Lambda K^+ K^-)=(15.9\pm 1.2\pm 1.2\pm 2.0)\times 10^{-6}$, which implies  the possibility for 
other contributions, such as that from the resonant $\Lambda_b\to K^- N^*,\,N^*\to\Lambda K^+$ decay with $N^*$ as a higher-wave baryon state. For $\Lambda_b\to \Lambda \eta^{(\prime)}$, we show that ${\cal B}(\Lambda_b\to \Lambda\eta,\,\Lambda\eta^\prime)=
(1.47\pm 0.35,1.83\pm 0.58)\times 10^{-6}$,  which are  consistent with the current data of $(9.3^{+7.3}_{-5.3},<3.1)\times 10^{-6}$, 
respectively. 
Our results also support  the relation of ${\cal B}(\Lambda_b\to \Lambda\eta) \simeq {\cal B}(\Lambda_b\to\Lambda\eta^\prime)$, 
given by the previous study.
\end{abstract}

\maketitle
\section{introduction}
The charmless two-body $\Lambda_b$ decays of $\Lambda_b\to p K^-$
and $\Lambda_b\to p\pi^-$ have been observed by the CDF
Collaboration~\cite{LbtopM} with the branching ratios of ${\it O}(10^{-6})$,
which are in accordance with the recent measurements
on $\Lambda_b\to \Lambda \phi$ and $\Lambda_b\to \Lambda
\eta^{(\prime)}$ by the LHCb Collaboration, given by~\cite{LbtoLphi,LbtoLeta}
\begin{eqnarray}\label{data1}
{\cal B}(\Lambda_b\to \Lambda \phi)&=&(5.18\pm 1.04\pm 0.35^{+0.67}_{-0.62})\times 10^{-6}\,,\nonumber\\
{\cal B}(\Lambda_b\to \Lambda \eta)&=&(9.3^{+7.3}_{-5.3})\times 10^{-6}\,,\nonumber\\
{\cal B}(\Lambda_b\to \Lambda \eta^{\prime})&<&3.1\times 10^{-6}\,,\;(\text{90\% C.L.})
\end{eqnarray}
where the evidence is seen for the $\eta$ mode at the level of 3$\sigma$-significance.

Theoretically,  $\Lambda_b\to p (K^{*-},\pi^-,\rho^-)$ decays
via $b\to u\bar u (d,s)$ at the quark level have been studied
in the literature~\cite{Lu:2009cm,Wang:2013upa,Wei:2009np,Hsiao:2014mua,Liu:2015qfa,Zhu:2016bra}.
In particular, it is interesting to point out that the direct CP violating asymmetry in
$\Lambda_b\to p K^{*-}$ is predicted to be as large as 20\%, which is promising to be observed
in the future measurements. On the other hand, the decay of
$\Lambda_b\to\Lambda \phi$ via $b\to s\bar s s$ 
has not been well explored even though both the decay branching ratio and 
T-odd triple-product asymmetries~\cite{Guo:1998eg,Arunagiri:2003gu,Leitner:2006nb}
 have been examined by the experiment at LHCb~\cite{LbtoLphi}.
According to the newly measured three-body $\Lambda_b\to\Lambda K^+ K^-$ decay by the LHCb Collaboration,
given by~\cite{LbtoLKK}
\begin{eqnarray}\label{data2}
{\cal B}(\Lambda_b\to\Lambda K^+ K^-)=(15.9\pm 1.2\pm 1.2\pm 2.0)\times 10^{-6}\,,
\end{eqnarray}
it implies a resonant $\Lambda_b\to\Lambda \phi,\,\phi\to K^+ K^-$ contribution
with the signal seen at the low range of $m^2(K^+ K^-)$ from the Dalitz plot.
However, to estimate this resonant contribution, one has to understand 
 ${\cal B}(\Lambda_b\to\Lambda \phi)$ in Eq.~(\ref{data1}) first.
Such a study is also important 
for further examinations  of the triple-product asymmetries~\cite{TPA}.
For $\Lambda_b\to \Lambda \eta^{(\prime)}$,
the relation of
${\cal B}(\Lambda_b\to \Lambda \eta)\simeq {\cal B}(\Lambda_b\to \Lambda \eta^{\prime})$
found in Ref.~\cite{Ahmady:2003jz} seems not to be consistent with the data in Eq.~(\ref{data1}).
Moreover, the first works on $\Lambda_b\to \Lambda \eta^{\prime}$ with the branching ratios predicted
to be $O(10^{-6}-10^{-5})$ in comparison with the data in Eq.~(\ref{data1}) were done
before the observations of $\Lambda_b\to p(K^-,\,\pi^-)$, which 
can be used to extract
the $\Lambda_b\to {\bf B}_n$ transition form factors from QCD models~\cite{Hsiao:2014mua,Gutsche:2014zna}.
For a reconciliation,
we would like  to reanalyze $\Lambda_b\to \Lambda \eta^{(\prime)}$.

In this work,
we will use the factorization approach for the theoretical calculations of
$\Lambda_b\to \Lambda\phi$ and $\Lambda_b\to\Lambda\eta^{(\prime)}$ 
as those in the $\Lambda_b\to p (K^{*-},\pi^-,\rho^-)$ decays~\cite{Hsiao:2014mua}.

\section{Formalism}
\begin{figure}[t!]
\centering
\includegraphics[width=2.5in]{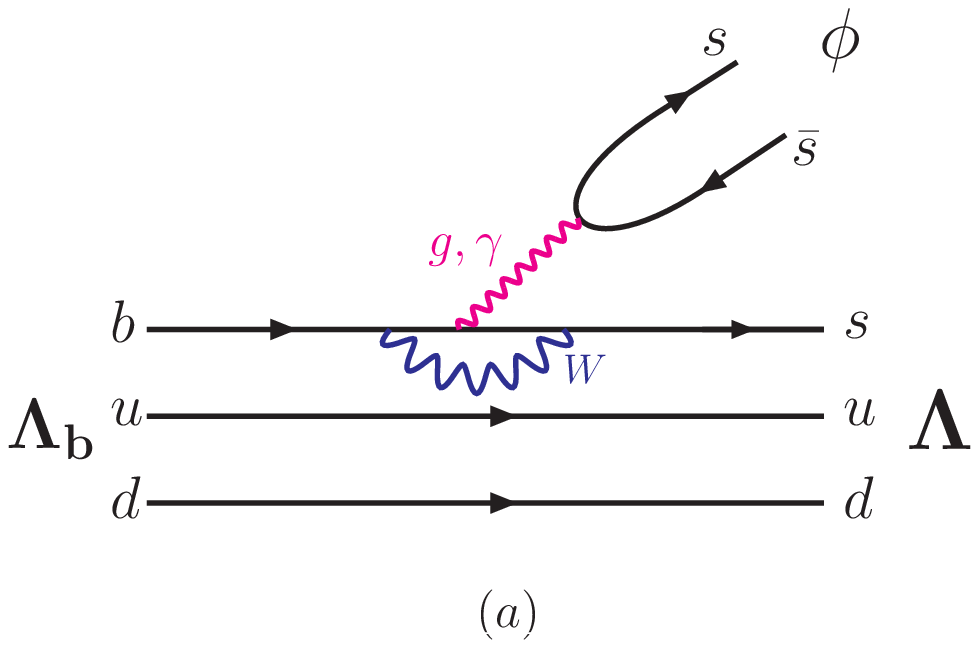}
\includegraphics[width=2.5in]{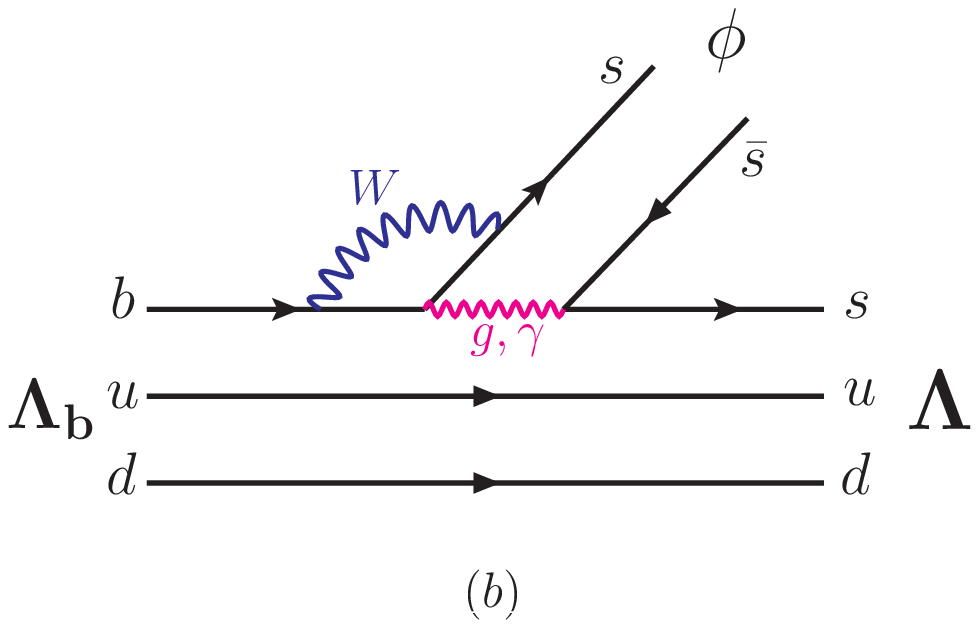}
\includegraphics[width=2in]{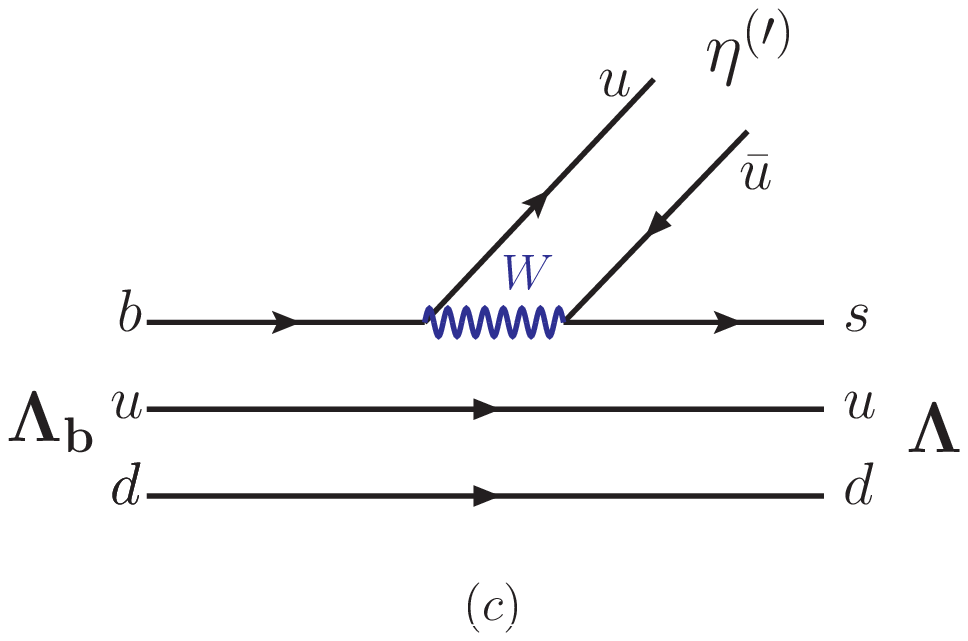}
\includegraphics[width=2in]{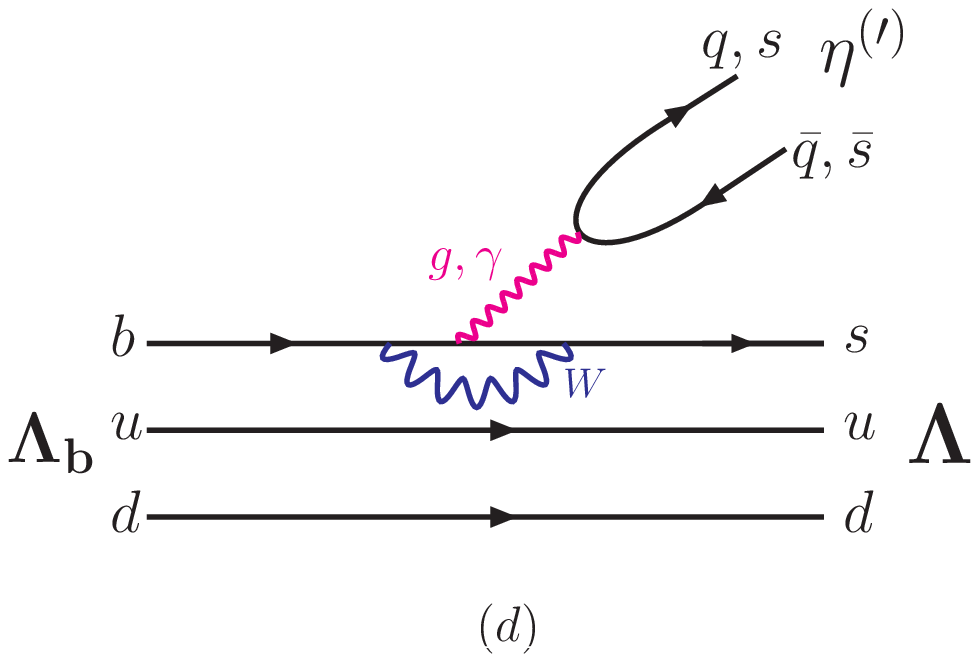}
\includegraphics[width=2in]{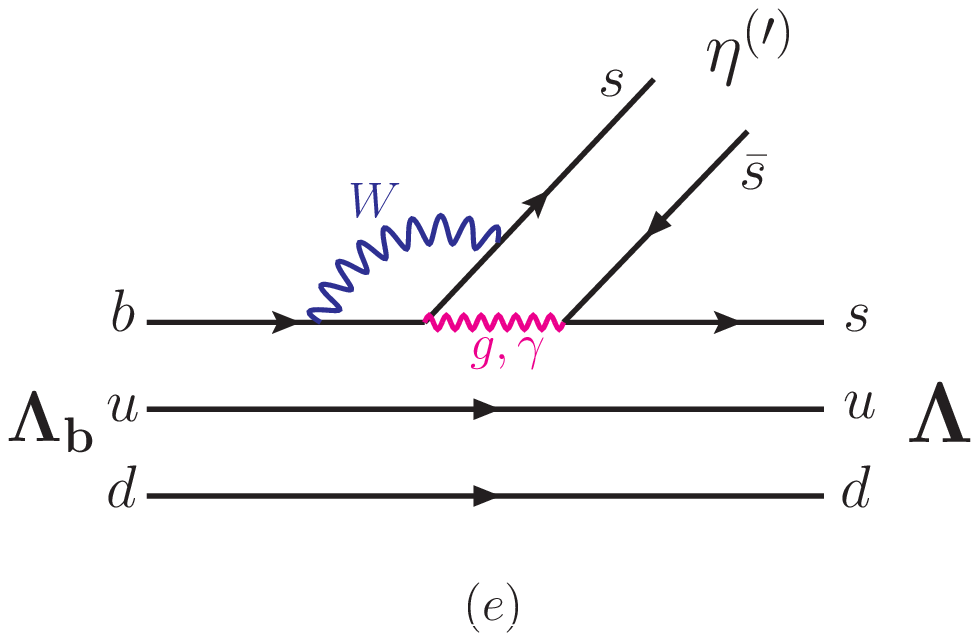}
\caption{Feynman diagrams (a,b) and (c,d,e)
from $\Lambda_b^0\to\Lambda\phi$ and 
$\Lambda_b\to \Lambda \eta^{(\prime)}$ decays, respectively.}\label{dia}
\end{figure}
In terms of the effective Hamiltonian for the charmless $b\to s s\bar s$ transition
at the quark level shown  Fig.~\ref{dia}, the amplitude of $\Lambda_b\to \Lambda \phi$
based on the factorization approach can be derived as~\cite{ali}
\begin{eqnarray}\label{amp1a}
&&{\cal A}(\Lambda_b\to \Lambda \phi)=\frac{G_F}{\sqrt 2}
\alpha_3\langle \phi|\bar s\gamma_\mu s|0\rangle
\langle \Lambda|\bar s\gamma_\mu(1-\gamma_5)b|\Lambda_b\rangle\,,
\end{eqnarray}
with $G_F$ the Fermi constant, $V_{q_1q_2}$
the Cabibbo-Kobayashi-Maskawa (CKM) matrix elements, and
$\alpha_3=-V_{tb}V_{ts}^*(a_3+a_4+a_5-a_9/2)$, where
$a_i\equiv c^{eff}_i+c^{eff}_{i\pm1}/N_c$ for $i=$odd (even)
are composed of the effective Wilson coefficients $c_i^{eff}$ defined in Ref.~\cite{ali}
with the color number $N_c$.
As depicted in Fig.~\ref{dia}, 
the amplitudes of $\Lambda_b\to \Lambda \eta^{(\prime)}$ are given by
\begin{eqnarray}\label{amp1b}
{\cal A}(\Lambda_b\to \Lambda \eta^{(\prime)})&=&
\frac{G_F}{\sqrt 2}\bigg\{
\bigg[\beta_2\langle \eta^{(\prime)}|\bar q\gamma_\mu\gamma_5 q|0\rangle+
\beta_3\langle \eta^{(\prime)}|\bar s \gamma_\mu \gamma_5 s|0\rangle\bigg]
\langle \Lambda|\bar s \gamma_\mu(1-\gamma_5) b|\Lambda_b\rangle\nonumber\\
&+&\beta_6
\langle \eta^{(\prime)}|\bar s\gamma_5 s|0\rangle\langle
\Lambda|\bar s(1-\gamma_5) b|\Lambda_b\rangle\bigg\}\,,
\end{eqnarray}
with $q=u$ or $d$, where $\beta_2=-V_{ub}V_{us}^*\,a_2+V_{tb}V_{ts}^*(2a_3-2a_5+a_9/2)$,
$\beta_3=V_{tb}V_{ts}^*(a_3+a_4-a_5-a_9/2)$, and $\beta_6=V_{tb}V_{ts}^*\,2a_6$.
The matrix elements of the $\Lambda_b\to \Lambda$ baryon
transition in Eqs. (\ref{amp1a}) and (\ref{amp1b}) have been
parameterized as~\cite{CF}
\begin{eqnarray}\label{ffs1}
\langle \Lambda|\bar s\gamma_\mu(1-\gamma_5)b|\Lambda_b\rangle&=&
\bar u_{\Lambda}(f_1\gamma_\mu-g_1\gamma_\mu \gamma_5)u_{\Lambda_b}\,,\nonumber\\
\langle \Lambda|\bar s(1-\gamma_5)b|\Lambda_b\rangle&=&
\bar u_{\Lambda}(f_S\gamma_\mu-g_P\gamma_\mu \gamma_5)u_{\Lambda_b}\,,
\end{eqnarray}
where $f_1$, $g_1$, $f_S$, and $g_P$ are the form factors, with
$f_S=[(m_{\Lambda_b}-m_{\Lambda})/(m_b-m_s)] f_1$ and
$g_P=[(m_{\Lambda_b}+m_{\Lambda})/(m_b+m_s)] g_1$
by virtue of equations of motion. 
Note that,
in Eq.~(\ref{ffs1}), we have neglected the form factors related to
$\bar u_{\Lambda}\sigma_{\mu\nu}q^\nu(\gamma_5)u_{\Lambda_b}$ and
$\bar u_{\Lambda}q_\mu (\gamma_5) u_{\Lambda_b}$ that flip the helicity~\cite{Gutsche:2013oea}.
With the double-pole momentum dependences,
$f_1$ and $g_1$  can be written as~\cite{Hsiao:2014mua}
\begin{eqnarray}
f_1(q^2)=\frac{f_1(0)}{(1-q^2/m_{\Lambda_b}^2)^2}\,,\;
g_1(q^2)=\frac{g_1(0)}{(1-q^2/m_{\Lambda_b}^2)^2}\,,
\end{eqnarray}
where we have taken $C_F(\Lambda_b\to \Lambda)\equiv f_1(0)=g_1(0)$ as the leading
approximation based on
the $SU(3)$ flavor and $SU(2)$ spin symmetries~\cite{Brodsky1,Hsiao:2015cda}.
We remark that the perturbative corrections to the  $\Lambda_b\to \Lambda$  transition form factors 
from QCD sum rules have been recently computed in Ref.~\cite{Wang_JHEP}.
Clearly, for more precise evaluations of the form factors,  these corrections
should be included.

The matrix elements in Eqs.~(\ref{amp1a}) and (\ref{amp1b})
for the meson productions read~\cite{Beneke:2002jn}
\begin{eqnarray}
&&\langle \phi|\bar s\gamma_\mu s|0\rangle=m_\phi f_\phi\varepsilon_\mu^*\,,
\langle \eta^{(\prime)}|\bar q\gamma_\mu \gamma_5 q|0\rangle=
-\frac{i}{\sqrt 2}f^q_{\eta^{(\prime)}} q_\mu\,,\nonumber\\
&&\langle \eta^{(\prime)}|\bar s\gamma_\mu \gamma_5 s|0\rangle=-if^s_{\eta^{(\prime)}} q_\mu\,,
2m_s \langle \eta^{(\prime)}|\bar s\gamma_5 s|0\rangle=-ih^s_{\eta^{(\prime)}}\,,
\label{decayconstant0}
\end{eqnarray}
with the polarization $\varepsilon_\mu^*$ and four-momentum $q_\mu$ vectors 
for $\phi$ and $\eta^{(\prime)}$, respectively, where
$f_\phi$, $f^q_{\eta^{(\prime)}}$, 
and $h^s_{\eta^{(\prime)}}$ are decay constants.
Unlike the usual decay constants, $f^q_{\eta^{(\prime)}}$ and $f^s_{\eta^{(\prime)}}$
are the consequences of the $\eta-\eta^\prime$ mixing, in which
the Feldmann, Kroll and Stech (FKS) scheme is adopted as~\cite{FKS}
\begin{eqnarray}\label{decayconstant}
\left(\begin{array}{c}
\eta\\
\eta'
\end{array}\right)
=
\left(\begin{array}{cc}
\cos\phi&-\sin\phi\\
\sin\phi&\cos\phi
\end{array}\right)
\left(\begin{array}{c}
\eta_q\\
\eta_s
\end{array}\right)\;,
\end{eqnarray}
with $|\eta_q\rangle=(|u\bar u+d\bar d\rangle)/\sqrt 2$ and $|\eta_s\rangle=|s\bar s\rangle$,
where the mixing angle is extracted as $\phi=(39.3\pm 1.0)^\circ$.
As a result, $f^q_{\eta^{(\prime)}}$ and $f^s_{\eta^{(\prime)}}$ actually mix with
the decay constants $f_q$ and $f_s$ for $\eta_q$ and $\eta_s$, respectively.
Note that $h^s_{\eta^{(\prime)}}$ in Eq.~(\ref{decayconstant0})  contains 
the contribution from the QCD anomaly, given by
\begin{eqnarray}
2m_s\langle \eta^{(\prime)}|\bar s i\gamma_5 s|0\rangle=
\partial^\mu \langle \eta^{(\prime)}|\bar s \gamma_\mu\gamma_5 s|0\rangle+
\langle \eta^{(\prime)}|\frac{\alpha_s}{4\pi}G\tilde{G}|0\rangle\,,
\end{eqnarray}
where $\alpha_s$ is the strong coupling constant, 
 $G(\tilde G)$ is the (duel) gluon field tensor,
 $\partial^\mu \langle \eta^{(\prime)}|\bar s \gamma_\mu\gamma_5 s|0\rangle
=f_{\eta^{(\prime)}}m_{\eta^{(\prime)}}^2$, and
$\langle \eta^{(\prime)}|\alpha_s G\tilde{G}|0\rangle \equiv 4\pi a_{\eta^{(\prime)}}$.
Explicitly, one has~\cite{Beneke:2002jn}
\begin{eqnarray}
h^s_{\eta^{(\prime)}}=a_{\eta^{(\prime)}}+f^s_{\eta^{(\prime)}} m_{\eta^{(\prime)}}^2\,,
\end{eqnarray}
which will be used in the numerical analysis.

\section{Numerical Results and Discussions}
For our numerical analysis,
the CKM matrix elements in the Wolfenstein parameterization are given by~\cite{pdg}
\begin{eqnarray}
(V_{ub},\,V_{us},\,V_{tb},\,V_{ts})=(A\lambda^3(\rho-i\eta),\lambda,1,-A\lambda^2)\,,
\end{eqnarray}
with $(\lambda,\,A,\,\rho,\,\eta)=(0.225,\,0.814,\,0.120\pm 0.022,\,0.362\pm 0.013)$.
\begin{table}[t]
\caption{$\alpha_i\,(\beta_i)$ with $N_c=2,\,3$, and $\infty$.}\label{alpha_i}
\begin{tabular}{|c|ccc|}
\hline
$\alpha_i\,(\beta_i)$ & $N_c=2$ &  $N_c=3$ & $N_c=\infty$ \\\hline
$10^4\alpha_3$ & $-21.97 - 4.47i$ &  $-15.51 - 3.39i$ & $-2.59 - 1.24i$ \\
$10^4 \beta_2$   & $-11.93 + 1.71i$ &  $-9.42+0.23i$ & $-4.41-2.73i$ \\
$10^4\beta_3$  & $7.58+3.18i$ &  $10.07+3.39i$ & $15.05+3.82i$ \\
$10^4\beta_6$ & $47.48+6.44i$ &  $49.55+6.87i$ & $53.71+7.73i$ \\
\hline
\end{tabular}
\end{table}
In Table~\ref{alpha_i}, we fix $N_c=3$ for $a_i$ but shift it from 2 to $\infty$
in the generalized version of the factorization approach
to take into account the non-factorizable effects as the uncertainty.
For the form factors, we use
$C_F(\Lambda_b\to \Lambda)=-\sqrt {2/3}\,C_F(\Lambda_b\to p)$~\cite{Hsiao:2015cda}
with $C_F(\Lambda_b\to p)=0.136\pm 0.009$~\cite{Hsiao:2014mua}.
Apart from $f_\phi=0.231$ GeV~\cite{Ball},
we adopt the decay constants for $\eta$ and $\eta^{\prime}$
from Ref.~\cite{Beneke:2002jn},  given by
\begin{eqnarray}\label{decay_const}
&&(f^q_{\eta},f^q_{\eta^\prime},f^s_{\eta},f^s_{\eta^\prime})=
(0.108,\,0.089\,,-0.111,\,0.136)\,\text{GeV}\,,\nonumber\\
&&(h^s_{\eta},h^s_{\eta^\prime})=(-0.055,\,0.068)\,\text{GeV}^3\,,
\end{eqnarray}
respectively. 
Subsequently, we obtain the branching ratios, given in Table~\ref{tab2}.
\begin{table}[b]
\caption{Numerical results for the branching ratios
with the first and second errors from the non-factorizable effects
and the form factors, respectively, in comparison with 
the experimental data~\cite{LbtoLphi,LbtoLeta}
and the study in Ref.~\cite{Ahmady:2003jz}.
Note that, in column 3, the two values without and with the parenthesis correspond to
the form factors in the approach of QCD sum rules and the pole model, respectively.}\label{tab2}
\begin{tabular}{|c|c|c|c|}
\hline
decay mode  &  our results & data~\cite{LbtoLphi,LbtoLeta}&Ref.~\cite{Ahmady:2003jz}\\\hline
$10^6{\cal B}\Lambda_b\to \Lambda\phi)$
&$1.77^{+1.76}_{-1.71}\pm 0.24$&$5.18\pm 1.29$&...\\
$10^6{\cal B}(\Lambda_b\to \Lambda\eta)$                     
&$1.47^{+0.29}_{-0.13}\pm 0.20$ &$9.3^{+7.3}_{-5.3}$&11.47\,(2.95)\\
$10^6{\cal B}(\Lambda_b\to \Lambda\eta^{\prime})$       
&$1.83^{+0.55}_{-0.18}\pm 0.25$ &$<3.1$&11.33\,(3.24)\\
\hline
\end{tabular}
\end{table}

As seen in Table~\ref{alpha_i}, $\alpha_3$ for  $\Lambda_b\to \Lambda\phi$
is sensitive to the non-factorizable effects.
In comparison with the data in Table~\ref{tab2} and Eq.~(\ref{data1}),
the $\Lambda_b\to \Lambda\phi$ decay is judged to receive the non-factorizable effects with $N_c=2$,
such that ${\cal B}(\Lambda_b\to \Lambda\phi)=(3.53\pm 0.24)\times 10^{-6}$.
With ${\cal B}(\phi\to K^+ K^-)=(48.5\pm 0.5)\%$~\cite{pdg}, we get
${\cal B}(\Lambda_b\to \Lambda(\phi\to) K^+ K^-)=(1.71\pm 0.12)\times 10^{-6}$,
which is much lower than
the data of $(15.9\pm 4.4)\times 10^{-6}$ in Eq.~(\ref{data2}),
leaving some room for other contributions, such as 
the resonant $\Lambda_b\to K^- N^*,\,N^*\to\Lambda K^+$ decay with
$N^*$ denoted as the higher-wave baryon state. 
Here, we would suggest
a more accurate experimental examination on the $\Lambda K$ invariant mass spectrum,
which depends on the peak around the threshold of $m_{\Lambda K}\simeq m_\Lambda+m_K$,
while the Dalitz plot might possibly reveal the signal~\cite{LbtoLKK}.
The result of ${\cal B}(\Lambda_b\to \Lambda\eta)=(1.47\pm 0.35)\times 10^{-6}$ 
in Table~\ref{tab2} shows a consistent result with the data due to its large
uncertainty.
On the other hand,
${\cal B}(\Lambda_b\to \Lambda\eta^\prime)=(1.83\pm 0.58)\times 10^{-6}$
agrees with the experimental upper bound.
The  $\eta$ and $\eta^\prime$ modes with ${\cal B}\simeq 10^{-6}$ are mainly resulted from
the form factor $C_F(\Lambda_b\to\Lambda)\sim 0.14$ extracted in Ref.~\cite{Hsiao:2014mua}
in agreement with the calculation in 
QCD models~\cite{CF,Wei:2009np,Gutsche:2013oea},
which explains why
${\cal B}(\Lambda_b\to p K^-,\,p \pi^-)$ are also around $10^{-6}$~\cite{pdg}.
As seen in Table~\ref{tab2}, 
our results for the $\eta^{(')}$ modes are smaller than those in Ref.~\cite{Ahmady:2003jz}.
In addition, we note that,
the result of
${\cal B}(\Lambda_b\to \Lambda\eta^\prime)=11.33(3.24)\times 10^{-6}$ 
in the QCD sum rule (pole) model~\cite{Ahmady:2003jz}, 
 apparently exceeds the data.
 However, the predictions for $ \Lambda_b\to \Lambda\eta$ in Ref.~\cite{Ahmady:2003jz}
 are still consistent with the current data.
We also point out that the relation of ${\cal B}(\Lambda_b\to \Lambda\eta)
\simeq {\cal B}(\Lambda_b\to\Lambda\eta^\prime)$
still holds as in Ref.~\cite{Ahmady:2003jz}.

It is known that the gluon content of  $\eta^{(\prime)}$ 
can contribute to the flavor-singlet $B\to K\eta^{(\prime)}$ decays in three ways~\cite{Beneke:2002jn}:
(i) the $b\to sgg$ amplitude related to the effective charm decay constant,
(ii) the spectator scattering involving two gluons, and (iii) the singlet weak annihilation. 
It is interesting to ask if these three production mechanisms are
 also relevant to the corresponding $\Lambda_b\to \Lambda\eta^{(\prime)}$ decays.
For (i), its contribution to $\Lambda_b\to \Lambda\eta^{(\prime)}$
has been demonstrated to be small~\cite{Ahmady:2003jz}
since,
by effectively relating $b\to sgg$ to $b\to s c\bar c$, 
the $c\bar c$ vacuum annihilation of $\eta^{(\prime)}$ is suppressed  due to
the decay constants 
$(f^c_{\eta},f^c_{\eta^{\prime}})\simeq (-1,-3)$ MeV~\cite{Beneke:2002jn} 
being much smaller than $f^q_{\eta^{(\prime)}}$ in Eq.~(\ref{decay_const}).
For (ii), since  one of the gluons from the spectator quark 
connects to the recoiled $\eta^{(\prime)}$, the contribution belongs to
 the non-factorizable effect, which has been inserted into the effective number of $N_c$ (from 2 to $\infty$)
 in our  generalized factorization approach.
For (iii), it is the sub-leading power contribution which does not
 contribute to $\Lambda_b\to\Lambda\eta^{(\prime)}$.

Finally, we remark that,
in the $b$-hadron decays, such as those of $B$ and $\Lambda_b$,
the generalized factorization with the floating $N_c=2 \rightarrow \infty$~\cite{ali}
can empirically estimate the non-factorizable effects, such that
it can be used to explain the data as well as make predictions. On the other hand,
the  QCD factorization~\cite{Beneke:2002jn} 
could in general calculate the non-factorizable effects in
some specific processes.
Although the current existing  studies on 
$\Lambda_b\to \Lambda (\phi,\eta^{(\prime)})$ are based on the generalized factorization,
it is useful to calculate  these decay modes in the QCD factorization.
In particular, since the decay of $\Lambda_b\to \Lambda\phi$ with $N_c=2$ has shown to be sensitive to
the non-factorizable effects,  its study in the QCD factorization is clearly interesting.

\section{Conclusions}
In sum, we have studied the charmless two-body 
$\Lambda_b\to \Lambda \phi$ and $\Lambda_b\to \Lambda \eta^{(\prime)}$ and three-body 
$\Lambda_b\to \Lambda K^+ K^-$
decays.
By predicting ${\cal B}(\Lambda_b\to \Lambda\phi)=(3.53\pm 0.24)\times 10^{-6}$
to agree with the observation, we have found that
${\cal B}(\Lambda_b\to \Lambda (\phi\to)K^+ K^-)=(1.71\pm 0.12)\times 10^{-6}$
cannot explain the observed 
${\cal B}(\Lambda_b\to \Lambda K^+ K^-)=(15.9\pm 4.4)\times 10^{-6}$, which 
leaves much room for the contribution from
the resonant $\Lambda_b\to K^- N^*,\,N^*\to\Lambda K^+$ decay.
We have obtained 
${\cal B}(\Lambda_b\to \Lambda\eta,\,\Lambda\eta^\prime)=
(1.47\pm 0.35,1.83\pm 0.58)\times 10^{-6}$
in comparison with the data of $(9.3^{+7.3}_{-5.3},<3.1)\times 10^{-6}$, 
respectively.
 In addition, our results still support the relation of ${\cal B}(\Lambda_b\to \Lambda\eta)
\simeq {\cal B}(\Lambda_b\to\Lambda\eta^\prime)$.
It is clear that future more precise experimental measurements on
the present $\Lambda_b$ decays are important to test the QCD models,
in particular the generalized factorization one.

\section*{ACKNOWLEDGMENTS}
The work was supported in part by
National Center for Theoretical Sciences,
National Science Council (NSC-101-2112-M-007-006-MY3) and
MoST (MoST-104-2112-M-007-003-MY3).


\begin{thebibliography}{99}
\bibitem{LbtopM}
T.~Aaltonen {\it et al.} [CDF Collaboration],
Phys.\ Rev.\ Lett.\  {\bf 103}, 031801 (2009).

\bibitem{LbtoLphi}
R.~Aaij {\it et al.} [LHCb Collaboration],
arXiv:1603.02870 [hep-ex].

\bibitem{LbtoLeta}
R.~Aaij {\it et al.} [LHCb Collaboration], JHEP {\bf 1509}, 006 (2015).


\bibitem{Lu:2009cm}
C.D.~Lu, Y.M.~Wang, H.~Zou, A.~Ali and G.~Kramer,
Phys.\ Rev.\ D {\bf 80}, 034011 (2009).

\bibitem{Wang:2013upa}
S.~Wang, J.~Huang and G.~Li,
  Chin.\ Phys.\ C {\bf 37}, 063103 (2013).

\bibitem{Wei:2009np}
Z.T.~Wei, H.W.~Ke and X.Q.~Li, Phys.\ Rev.\ D {\bf 80}, 094016 (2009).

\bibitem{Zhu:2016bra}
J.~Zhu, H.W.~Ke and Z.T.~Wei,
arXiv:1603.02800 [hep-ph].

\bibitem{Hsiao:2014mua}
Y.K.~Hsiao and C.Q.~Geng,
Phys.\ Rev.\ D {\bf 91}, 116007 (2015);
PoS FPCP {\bf 2015}, 073 (2015).

\bibitem{Liu:2015qfa}
Y.~Liu, X.H.~Guo and C.~Wang, Phys.\ Rev.\ D {\bf 91}, 016006 (2015).


\bibitem{Guo:1998eg}
X.H.~Guo and A.W.~Thomas,
Phys.\ Rev.\ D {\bf 58}, 096013 (1998).

\bibitem{Arunagiri:2003gu}
S.~Arunagiri and C.Q.~Geng,
Phys.\ Rev.\ D {\bf 69}, 017901 (2004).

\bibitem{Leitner:2006nb}
O.~Leitner, Z.J.~Ajaltouni and E.~Conte,
hep-ph/0602043.


\bibitem{LbtoLKK}
R.~Aaij {\it et al.} [LHCb Collaboration], arXiv:1603.00413 [hep-ex].

\bibitem{TPA}
C.H.~Chen and C.Q.~Geng,
Phys.\ Rev.\ D {\bf 63}, 054005 (2001);  Phys.\ Rev.\ D {\bf 63}, 114024 (2001);
Phys.\ Rev.\ D {\bf 64}, 074001 (2001); Phys.\ Rev.\ D {\bf 65}, 091502 (2002).
 


\bibitem{Ahmady:2003jz}
M.R.~Ahmady, C.S.~Kim, S.~Oh and C.~Yu,
Phys.\ Lett.\ B {\bf 598}, 203 (2004).

\bibitem{Gutsche:2014zna}
T.~Gutsche,
M.A.~Ivanov, J.G.~Kšrner, V.E.~Lyubovitskij and P.~Santorelli,
Phys.\ Rev.\ D {\bf 90}, 114033 (2014).



\bibitem{ali} A. Ali, G. Kramer and C.D. Lu, Phys. Rev.  D{\bf 58}, 094009 (1998).

\bibitem{CF}
A.~Khodjamirian, C.~Klein, T.~Mannel and Y.M.~Wang,
JHEP {\bf 1109}, 106 (2011); T.~Mannel and Y.M.~Wang,
JHEP {\bf 1112}, 067 (2011).

\bibitem{Gutsche:2013oea}
T.~Gutsche, M.A.~Ivanov, J.G.~Kšrner, V.E.~Lyubovitskij and P.~Santorelli,
Phys.\ Rev.\ D {\bf 88}, 114018 (2013).



\bibitem{Brodsky1}
G.P.~Lepage and S.J.~Brodsky, Phys.\ Rev.\ Lett.\  {\bf 43}, 545(1979) [Erratum-ibid.\  {\bf 43}, 1625 (1979)].

\bibitem{Hsiao:2015cda}
Y.K.~Hsiao, P.Y.~Lin, C.C.~Lih and C.Q.~Geng,
Phys.\ Rev.\ D {\bf 92}, 114013 (2015).


\bibitem{Wang_JHEP}
Y.M.~Wang and Y.L.~Shen, JHEP {\bf 1602}, 179 (2016).
  
\bibitem{Beneke:2002jn}
M.~Beneke and M.~Neubert,
Nucl.\ Phys.\ B {\bf 651}, 225 (2003).


\bibitem{FKS}
T.~Feldmann, P.~Kroll and B.~Stech,
Phys.\ Rev.\ D {\bf 58}, 114006 (1998); Phys.\ Lett.\ B {\bf 449}, 339 (1999).

\bibitem{pdg}
K.A.~Olive {\it et al.}  [Particle Data Group Collaboration], Chin.\ Phys.\ C {\bf 38}, 090001 (2014).

\bibitem{Ball} P.~Ball and R.~Zwicky, Phys.\ Rev.\ D {\bf 71}, 014029 (2005).

\end{thebibliography}
\end{document}